\documentstyle[aps,multicol,epsfig,amsmath,prb]{revtex}

\begin{document}

\date{May 17, 2000}

\title{Exploiting exciton-exciton interactions in
       semiconductor quantum dots for
       quantum-information processing}

\author{Filippo Troiani, Ulrich Hohenester, and Elisa Molinari} 

\address{Istituto Nazionale per la Fisica della Materia (INFM) and\\
Dipartimento di Fisica, Universit\`a di Modena e Reggio Emilia, 
Via Campi 213/A, 41100 Modena, Italy}

\maketitle

\begin{abstract} 

We propose an all-optical implementation of quantum-information
processing in semiconductor quantum dots, where electron-hole
excitations (excitons) serve as the computational degrees of freedom
(qubits). We show that the strong dot confinement leads to an overall
enhancement of Coulomb correlations and to a strong renormalization of
the excitonic states, which can be exploited for performing conditional
and unconditional qubit operations.


\end{abstract}

\pacs{85.30.Vw,03.67.-a,71.35.-y,73.20.Dx}


\begin{multicols}{2}


Coherent-carrier control in semiconductor nanostructures has recently
attracted an enormous interest~\cite{heberle:95,bonadeo:98,poetz:99}
because it allows the coherent manipulation of carrier wavefunctions on
a timescale shorter than typical dephasing times. Such a controlled
``wavefunction engineering'' is an essential prerequisite for a
successful implementation of ultrafast optical switching and
quantum-information processing in a pure solid state system. In this
respect, semiconductor quantum dots (QDs) appear to be among the most
promising candidates since the strong quantum confinement gives rise to
a discrete atomic-like density of states and, in turn, to a suppressed
coupling to the solid-state environment (e.g.,
phonons).~\cite{hawrylak:98,bimberg:98} It thus has been envisioned
that optical excitations in QDs could be successfully exploited for
quantum-information processing.~\cite{zanardi:98,quiroga:99}

Recently, optical spectroscopy of single QDs has revealed a
surprisingly rich
fine-structure:~\cite{motohisa:98,landin:98,dekel:98,zrenner:98}  With
increasing excitation density additional emission peaks appear in the
spectra, which supersede the original lines at even higher densities.
It has been demonstrated that these optical nonlinearities are due to
the strong quantum confinement, which results in a strongly reduced
phase space and an overall enhancement of Coulomb
correlations.~\cite{hohenester:99a,hawrylak:99} In fact, whenever one
additional electron-hole pair (exciton) is added to the QD, the optical
spectrum must change because of the resulting additional
exciton-exciton interactions.

In this Rapid Communication we show that such optical nonlinearities in
semiconductor QDs can be successfully exploited for performing
elementary quantum-computational operations (quantum gates). More
specifically, we shall identify the different excitonic states in QDs
with the computational degrees of freedom (qubits), and we will show
for a prototypical dot structure that a coherent manipulation of
biexcitonic resonances can perform conditional two-qubit operations.
Indeed, such controlled-NOT (C-NOT) operations, which form a
cornerstone for any implementation of quantum-information processing,
naturally arise from the strong internal interaction channels in the
electron-hole system.  To complete the requirements for implementing a
universal set of quantum gates, we will demonstrate that single qubits
can be manipulated by performing two C-NOT operations at different
photon energies. Our model for calculating the correlated electron-hole
states in the dot is based on realistic material and dot parameters;
this will allow us to argue that an implementation of the proposed
scheme should be possible with the present state-of-the-art sample
growth and coherent-carrier control.


The initial ingredients of our calculations are the single-particle
states, which we numerically obtain for a prototypical dot confinement
that is parabolic in the $(x,y)$-plane and box-like along $z$; such
confinement potentials have been demonstrated to be a particularly good
approximation for various kinds of self-assembled
dots.~\cite{hawrylak:98,rinaldi:96} We compute the single-particle
states $|\mu_{e,h}\rangle$ and energies $\epsilon_{\mu_{e,h}}^{e,h}$
for electrons and holes by numerically solving the three-dimensional
single-particle Schr\"odinger equation within the envelope-function and
effective-mass approximations.~\cite{dot}

Let us suppose that the dot is subjected to a series of short laser
pulses. We describe the light-matter coupling within the usual dipole
and rotating-wave approximations:

\begin{equation}\label{eq:op}
  {\cal H}_{\rm op}=-\frac 1 2
  \sum_i{\cal E}_o^{(i)}(t)\bigl\lgroup
  e^{i\omega^{(i)} t}\hat P+e^{-i\omega^{(i)} t}\hat P^\dagger\bigr\rgroup,
\end{equation}

\noindent where ${\cal E}_o^{(i)}(t)$ is the time envelope of the
$i$-th laser pulse with central frequency $\omega^{(i)}$, and $\hat
P=\sum_{\mu_e,\nu_h}M_{\mu_e\nu_h}^* d_{\nu_h}c_{\mu_e}$ is the
interband-polarization operator, with $M_{\mu_e\nu_h}$ the
dipole-matrix element for the optical transition between $\mu_e$ and
$\nu_h$; the field operators $c_{\mu_e}^\dagger$ ($d_{\nu_h}^\dagger$)
create an electron in state $\mu_e$ (hole in state $\nu_h$). To
illustrate the action of ${\cal H}_{\rm op}$ on the quantum dot, let us
assume that before the first laser pulse the dot is in its groundstate
$|{\rm vac}\rangle$ (i.e., no electrons in the conduction band and no
holes in the valence band). On arrival of the first pulse, the laser
light will create (through $\hat P^\dagger$) an interband polarization,
i.e., a superposition between the vacuum state and the different
excitonic states, which will propagate in phase with the driving
laser.  It is precisely this coherent evolution of the interband
polarization, which then allows the subsequent pulses to coherently
manipulate the correlated QD electron-hole states and to perform
specific quantum-computational operations.

As an important step within our proposal, we use the fact that in most
semiconductors electron-hole pairs with given spin orientation can be
selectively created by photons with a well-defined circular
polarization. Throughout this paper, we shall only consider excitons
with parallel spin orientations because of their strongly reduced
available phase space and the resulting simplified optical density of
states.~\cite{hohenester:99a} Moreover, as will be discussed further
below, within the proposed scheme we can restrict ourselves to single
excitons and biexcitons. We expand the exciton and biexciton states
within the subspaces of spin-selective electron-hole excitations, i.e.,
$\Psi_{\mu_e}^{\nu_h;\;x}$ and
$\Psi_{\mu_e\mu_e'}^{\nu_h\nu_h';\;\lambda}$ (the latter states are
antisymmetric with respect to the exchange of the two electron and hole
coordinates). The excitonic eigenenergies and eigenstates are then
obtained from the solutions of the two- and four-particle Schr\"odinger
equations:

\end{multicols}

\begin{mathletters}\label{eq:exciton.states}
\begin{eqnarray}\label{eq:exciton}
  E_x \Psi_{\mu_e}^{\nu_h;\;x}&=&
  (\epsilon_{\mu_e}^e+\epsilon_{\nu_h}^h)\Psi_{\mu_e}^{\nu_h;\;x}
  +\langle\mu_e,\nu_h|{\cal H}^{eh}|
  \bar\mu_e,\bar\nu_h\rangle
  \Psi_{\bar\mu_e}^{\bar\nu_h;\;x}\\
  \label{eq:biexciton}
  E_\lambda\Psi_{\mu_e\mu_e'}^{\nu_h\nu_h';\;\lambda}&=&
  (\epsilon_{\mu_e}^e+\epsilon_{\mu_e'}^e+
   \epsilon_{\nu_h}^h+\epsilon_{\nu_h'}^h)
   \Psi_{\mu_e\mu_e'}^{\nu_h\nu_h';\;\lambda}\nonumber\\&&
   +\langle \mu_e,\mu_e'|{\cal H}^{ee}|\bar\mu_e,\bar\mu_e'\rangle
      \Psi_{\bar\mu_e\bar\mu_e'}^{\nu_h\nu_h';\;\lambda}
   +\langle \nu_h,\nu_h'|{\cal H}^{hh}|\bar\nu_h,\bar\nu_h'\rangle
      \Psi_{\mu_e\mu_e'}^{\bar\nu_h\bar\nu_h';\;\lambda}\nonumber\\&&
   +\langle \mu_e,\nu_h|{\cal H}^{eh}|\bar\mu_e,\bar\nu_h\rangle
    \Psi_{\bar\mu_e\mu_e'}^{\bar\nu_h\nu_h';\;\lambda}
   +\langle \mu_e,\nu_h'|{\cal H}^{eh}|\bar\mu_e,\bar\nu_h'\rangle
    \Psi_{\bar\mu_e\mu_e'}^{\nu_h\bar\nu_h';\;\lambda}\nonumber\\&&
   +\langle \mu_e',\nu_h|{\cal H}^{eh}|\bar\mu_e',\bar\nu_h\rangle
    \Psi_{\mu_e\bar\mu_e'}^{\bar\nu_h\nu_h';\;\lambda}
   +\langle \mu_e',\nu_h'|{\cal H}^{eh}|\bar\mu_e',\bar\nu_h'\rangle
    \Psi_{\mu_e\bar\mu_e'}^{\nu_h\bar\nu_h';\;\lambda},
\end{eqnarray}
\end{mathletters}

\begin{multicols}{2}

\noindent with the Coulomb matrix elements accounting for
electron-electron (${\cal H}^{ee}$), hole-hole (${\cal H}^{hh}$), and
electron-hole (${\cal H}^{eh}$) interactions; in
Eq.~(\ref{eq:exciton.states}) an implicit summation over $\bar\mu$ and
$\bar\nu$ has been assumed. In our computational approach, we keep for
electrons and holes, respectively, the 10 energetically lowest
single-particle states and solve Eqs.~(\ref{eq:exciton.states}a,b) by
direct diagonalization of the Hamiltonian matrix; this approach was
recently proven successful in giving a realistic description of
experiment.~\cite{hohenester:00}


Fig.~1(a) shows the linear absorption spectrum of the dot~\cite{dot} as
computed from the solutions of
Eq.~(\ref{eq:exciton}).~\cite{absorption} We observe two pronounced
absorption peaks ($X_0$ and $X_1$) with an energy splitting of the
order of the confinement energy. A closer inspection of the exciton
wavefunctions $\Psi_{\mu_e}^{\nu_h;\;x}$ reveals that: the dominant
contribution of the groundstate exciton $X_0$ is from the energetically
lowest electron and hole single-particle states (with $s$-type
character); the dominant contribution of the $X_1$ exciton is from the
first excited electron and hole single-particle states (with $p$-type
character). This $p$-shell, in general, is two-fold degenerate and can
host two excitons with parallel spin orientation; however, because of
Coulomb coupling only one of these two excitons couples to the light
field (for a more detailed discussion see, e.g.,
Refs.~\onlinecite{hohenester:99a,hawrylak:99,hohenester:99b}).

Fig.~1(b) shows the absorption spectrum for the dot initially prepared
in the exciton groundstate $|X_0\rangle$. Because of state filling, the
character of the corresponding optical transition changes from
absorption to gain (i.e., negative absorption). Moreover, the
higher-energetic transition is shifted to lower energy; this pronounced
redshift is attributed to the formation of a biexcitonic state, whose
energy is reduced by an amount of $\Delta\approx 8$ meV because of
exchange interactions between the two electrons and holes,
respectively.~\cite{hawrylak:99} A closer inspection of the
biexcitonic wavefunction shows that the dominant contribution is from
excitons $X_0$ and $X_1$, and we therefore label this state with
$|X_0+X_1\rangle$.  Finally, Figs.~1(c,d) report the absorption spectra
for the dot initially prepared in state $X_1$ and $X_0+X_1$,
respectively. We checked that optical excitations within the energy
region shown in Fig.~1 can not excite any higher-excitonic states.


As the central step within the present proposal, we next assign
according to Table I the excitonic states to the computational degrees
of freedom (qubits). In the following, we shall demonstrate that for
this choice it is indeed possible to perform the conditional and
unconditional qubit operations required for quantum-information
processing. We first observe in Fig.~1 that the appearance and
disappearance of peaks at the frequencies indicated by the (solid and
dashed) shaded areas {\em conditionally}\/ depends on the setting of
specific qubits: E.g., the optical transitions at $\omega_{X_0}-\Delta$
is {\em only}\/ present if the second qubit is set equal to one
(Figs.~1(c) and 1(d)), whereas the transition at $\omega_{X_0}$ {\em
only}\/ appears if the second qubit is set equal to zero (Figs.~1(a)
and 1(b)); an analogous behaviour can be found for the other two
transitions. Indeed, it is this conditional on- and off-switching of
optical transitions that enables $\pi/2$-laserpulses to modify the
state of one qubit or not, depending on the setting of the other
qubit.

To illustrate the essence of our scheme, in the following we consider
the somewhat simplified excitation scenario of a laser pulse with
rectangular-shaped envelope ${\cal E}_o(t)$ and with central frequency
$\omega_{X_0}-\Delta$, and we assume that only transitions at
$\omega_{X_0}-\Delta$ are affected by the laser (results of our more
realistic simulations will be presented 

\centerline{\psfig{file=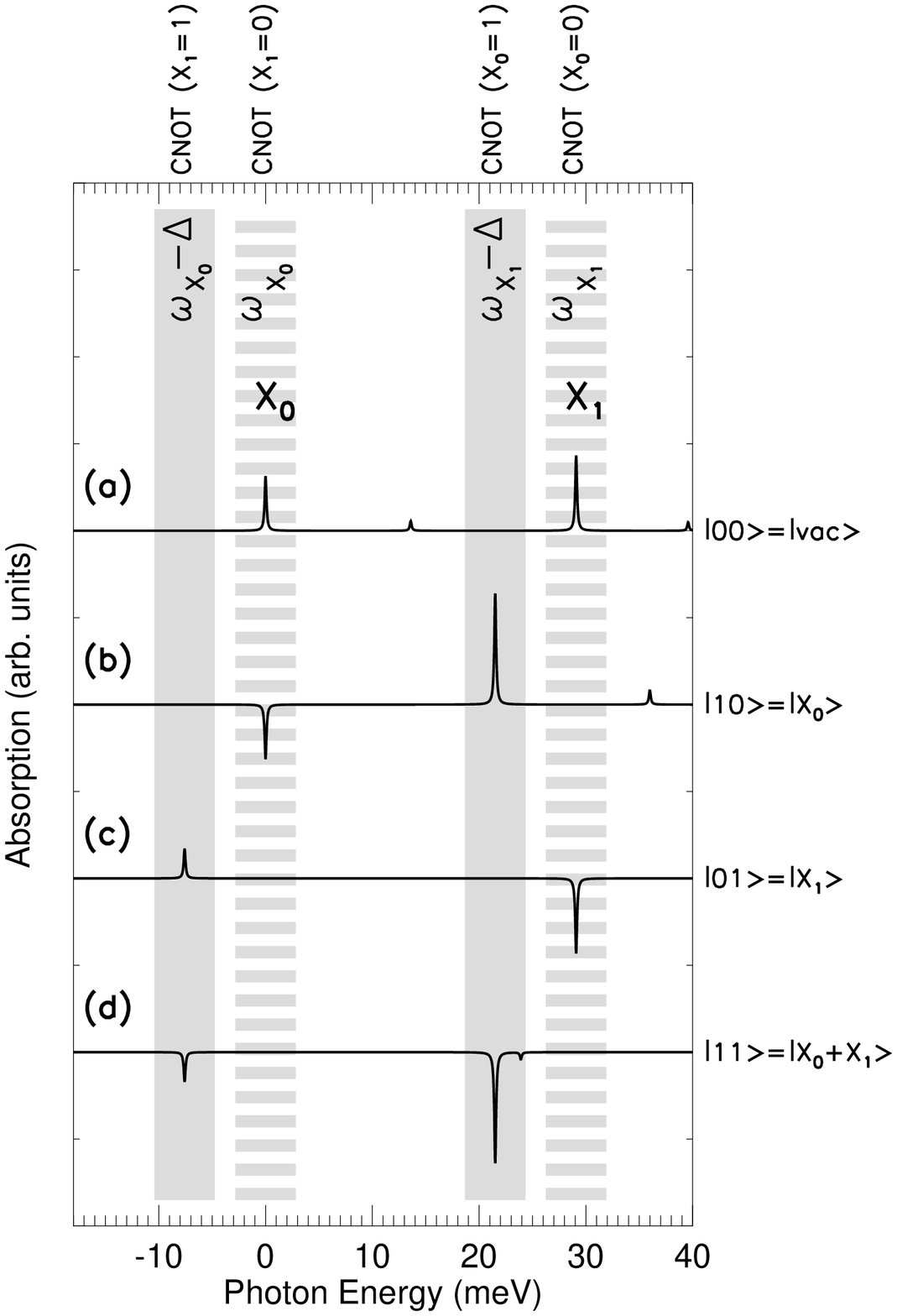,width=3.2in}}

{\small FIG. 1. Absorption spectra for semiconductor quantum dot
described in the text, which is initially prepared in: (a) vacuum
state; (b) exciton $|X_0\rangle$ state (exciton groundstate); (c)
exciton $|X_1\rangle$ state; (d) biexciton $|X_0+X_1\rangle$ state.
Photon energy zero is given by the groundstate exciton
$X_0$.~\cite{absorption}}

\noindent further below). Then, the
effective qubit-light Hamiltonian of Eq.~(\ref{eq:op}) reads:

\centerline{\psfig{file=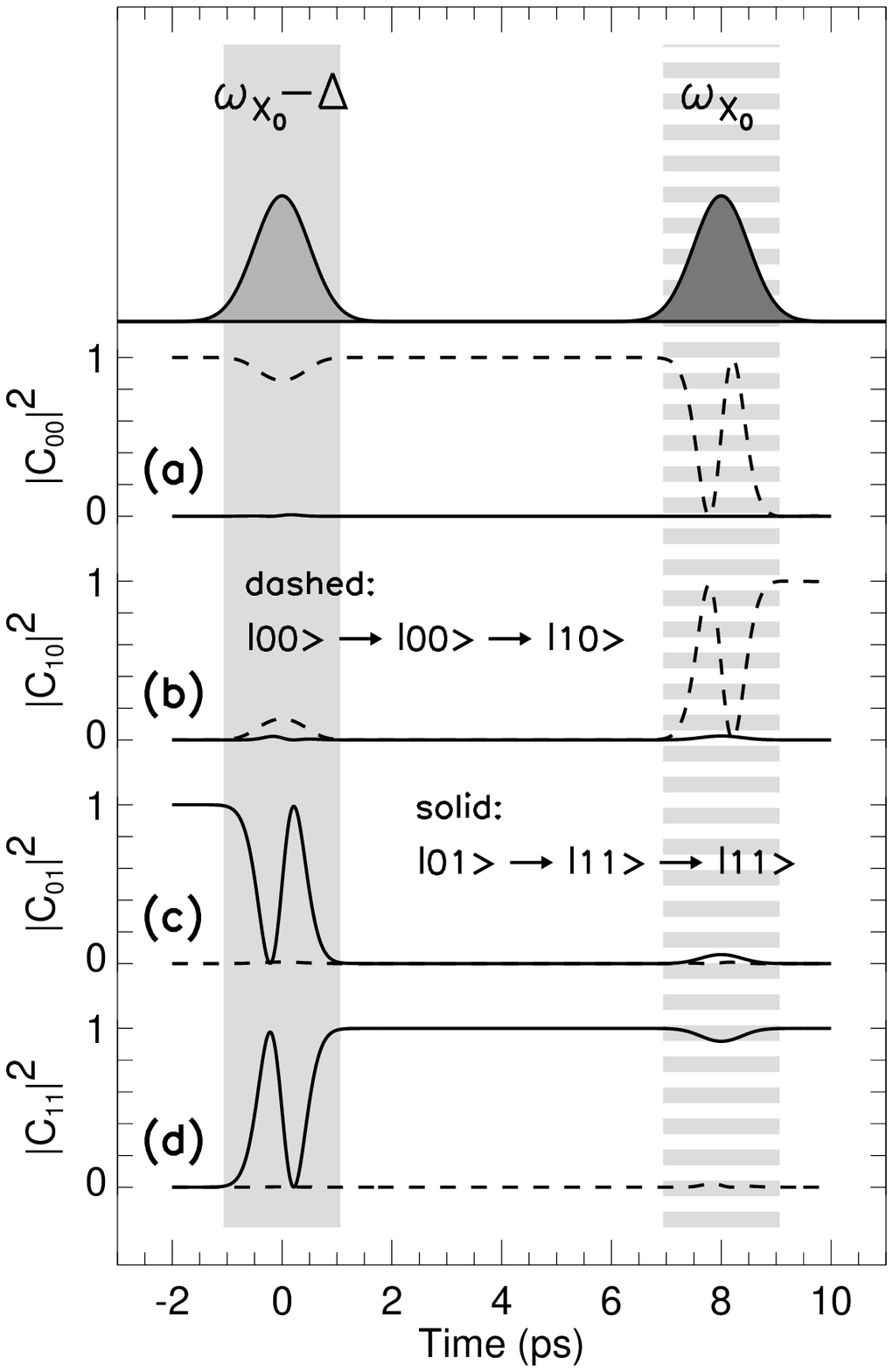,width=3.0in}}

{\small FIG. 2. Results of our simulations (neglecting dephasing) of
qubit manipulations by means of coherent-carrier control. Here,
$c_{00}$, $c_{10}$, $c_{01}$ and $c_{11}$ are the coefficients of the
state vector corresponding to the different qubit configurations. The
solid (dashed) lines present results of simulations where the system is
initially prepared in the $|01\rangle$ ($|00\rangle$) state. The
sequence of pulses and the corresponding photon energies are indicated
at the top of the figure; for the envelopes of the laser pulses we use
Gaussians $\propto\exp(-t^2/2\tau^2)$, with $\tau=0.5$ ps. The first
operation at time zero corresponds to a C-NOT operation where the
second qubit acts as the control-qubit; the sequence of the two pulses
corresponds to a NOT operation on the first qubit.}

\end{multicols}

\begin{equation}
  {\cal H}_{\rm op}=- \frac 1 2
  M{\cal E}_o(t)\bigl\lgroup
  e^{ i\omega t}|11\rangle\langle 01|+
  e^{-i\omega t}|01\rangle\langle 11|
  \bigr\rgroup_{\omega=\omega_{X_0}-\Delta},
\end{equation}

\noindent with the dipole-matrix element $M=\langle X_0|\hat P
|X_0+X_1\rangle$. If the duration of the pulse is $T$, its action on the
system can be expressed through the unitary transformation:

\begin{eqnarray}\label{eq:cnot}
 {\cal U}(t>T,0) = \left( \begin{array}{cccc}
   1 & 0 & 0 & 0 \\
   0 & e^{-iE_{X_0}t}  & 0 & 0 \\
   0 & 0 & \phantom{-i^{_{X_0}}}e^{-iE_{X_1}t}\cos(\Omega_RT) &
                            -i e^{-iE_{X_1}t}\sin(\Omega_RT)\\
   0 & 0 & -ie^{-iE_{X_0+X_1}t} \sin(\Omega_RT) & 
                 e^{-iE_{X_0+X_1}t} \cos(\Omega_RT)\\ 
   \end{array}\right),
\end{eqnarray}

\noindent which is supposed to act on a state vector with components
$(|00\rangle,|10\rangle,|01\rangle,|11\rangle)$; here $\Omega_R=M{\cal
E}_o$ is the Rabi frequency. Apparently, ${\cal U}(t,0)$ manipulates
the first qubit {\em only}\/ if the second qubit is set equal to one.
The details of 

\begin{multicols}{2}

\noindent such manipulations depend on the specific choices for
$t$ and $T$; e.g., for $\Omega_RT\;\text{mod}\;2\pi=\pm \frac\pi 2$
(i.e., $\pi/2$-pulse), $E_{X_0}t\;\text{mod}\;2\pi=0$,
$E_{X_1}t\;\text{mod}\;2\pi=\pm \frac\pi 2$, and
$E_{X_0+X_1}t\;\text{mod}\;2\pi=\pm \frac\pi 2$, the unitary
transformation of Eq.~(\ref{eq:cnot}) precisely corresponds to a
controlled NOT (C-NOT) operation.~\cite{remark:commensurable} By
inspection of Fig.~1, we observe that a completely analogous scheme,
although with different conditional qubit settings, applies for the
optical transitions at different photon energies (the conditional
dependence is indicated at the top of Fig.~1). Within our proposed
scheme, the unconditioned qubit operations (i.e., rotation of a single
qubit), which are requested besides the C-NOT operations for an
implementation of an universal set of gates, can be simply achieved by
combining two conditional operations at different photon energies:
E.g., in order to perform a NOT operation on the first qubit
(independent of the setting of the second qubit) we first have to
perform a C-NOT operation at frequency $\omega_{X_0}-\Delta$, followed
by a C-NOT operation at $\omega_{X_0}$ (see also Fig.~2; other
single-qubit rotations can be performed by choosing different durations
$T$ of the laserpulses); in principle, it is not even necessary to
perform the two operations in sequence, but one could use a single
appropriately tuned two-color laserpulse instead.

\vspace*{0.2cm}

{\small TABLE I. Assignment of the excitonic states to the
computational degrees of freedom (qubits).}

\noindent
\begin{tabular}{llc}
\tableline\tableline
Excitonic state       &  Energy                      &   Qubit state   \\ 
\tableline
$|{\rm vac}\rangle$   &  $0$                         &   $|00\rangle$  \\
$|X_0\rangle$         &  $E_{X_0}$                   &   $|10\rangle$  \\
$|X_1\rangle$         &  $E_{X_1}$                   &   $|01\rangle$  \\
$|X_0+X_1\rangle$     &  $E_{X_0+X_1}=E_{X_0}+E_{X_1}-\Delta$  
                                                     &   $|11\rangle$  \\
\tableline\tableline
\end{tabular}

\vspace*{0.4cm}

We finally address the possibility of experimental realization of our
proposed scheme. Quite generally, the feasibility of
quantum-information processing depends on the number of qubit
operations which can be performed on the time scale of dephasing. We
therefore have to estimate the time required for performing single
qubit manipulations (i.e., the duration $T_{\rm pulse}$ of the laser
pulse) and the typical dephasing times ($T_{\rm dephasing}$). In order
to perform within our scheme qubit manipulations at certain frequencies
without affecting the other transitions, it is necessary that the
spectral width of the laser pulse is narrow as compared to the energy
separation between neighbour peaks; from the basic relation
$\Delta\times T_{\rm pulse}\sim\hbar$ we can estimate a lower limit of
$T_{\rm pulse}\gtrsim 0.1$ ps (note that much longer pulses would be
required in case of non-spin-selective excitations, where the typical
energy splitting of peaks in the optical spectra would be of the order
of 1 meV). We also performed for the studied dot structure simulations
of qubit manipulations (C-NOT and unconditioned qubit rotations).
Figure 2 shows typical results of our simulations; for conceptual
clarity, we have chosen rather strong laser pulses to demonstrate that
even for such high electric fields the laser only manipulates
components of the state vector with appropriate energies.  From our
simulations we safely conclude that manipulation times $T_{\rm
pulse}\approx 0.25$ ps completely suffice to suppress any population of
components of the state vector at non-matching energies. For the
typical dephasing times of optical excitations in semiconductor QDs, we
assume a conservative estimate of $T_{\rm dephasing}\approx 40$ ps,
which is based on the results of a recent ingenious experiment of
Bonadeo et al.;~\cite{bonadeo:98} thus, $T_{\rm pulse}\ll T_{\rm
dephasing}$. Since the primary requirement for a successful
implementation of the proposed scheme is the coherent optical
manipulation of single excitons and biexcitons, we strongly believe
that an experimental realization is possible within the superb
standards of presentday coherent-carrier control. We finally stress
that one of the major advantages of our proposed scheme is the fact
that quantum gates can be solely performed by means of ultrashort laser
pulses and no additional external fields, which would introduce longer
manipulation times and additional decoherence, are required.


In conclusion, we have presented an all-optical implementation of
quantum-information processing in semiconductor QDs. Identifying the
qubits with excitonic states, we have demonstrated that both
conditional and unconditional qubit operations can be performed. We
have discussed that the implementation of the proposed scheme is
experimentally possible with the available tools of sample growth and
coherent-carrier control. We expect the present proposal to be
particularly promising for the first successful demonstration of
quantum-information processing in this class of materials.
Generalization of our scheme to three or four qubits (i.e., three or
four excitons) is possible, where further scaling is not easy because
of the more complicated optical density-of-states and the resulting
difficulties to make frequencies commensurable. In this respect, sample
quality and the question of how many excitons can be hosted in a single
dot without suffering losses due to Auger-type processes will be of
essential importance. Future work will also address excitonic states in
arrays of coupled semiconductor QDs, where similar excitonic
renormalization effects would allow the implementation of
quantum-information processing for a moderate number of qubits.

We are grateful to F. Rossi for most helpful discussions.  This work
was supported in part by the EU under the TMR Network "Ultrafast" and
the IST Project SQID, and by INFM through grant PRA-SSQI. U.H.
acknowledges support by the EC through a TMR Marie Curie Grant.

\end{multicols}


\begin{thebibliography}{10}


\bibitem{heberle:95}
A.P. Heberle, J.J. Baumberg, and K. K\"ohler, 
Phys. Rev. Lett. {\bf 75}, 2598 (1995).

\bibitem{bonadeo:98}
N.H. Bonadeo, J. Erland, D. Gammon, D.S. Katzer, D. Park, and D.G. Steel,
Science {\bf 282}, 1473 (1998).

\bibitem{poetz:99}
X. Hu and W. P\"otz, Phys. Rev. Lett. {\bf 82}, 3116 (1999);
P\"otz, Phys. Rev. Lett. {\bf 79}, 3262 (1997).

\bibitem{hawrylak:98}
L. Jacak, P. Hawrylak, and A. Wojs, 
{\em Quantum Dots}\/ (Springer, Berlin, 1998).

\bibitem{bimberg:98}
D. Bimberg, M. Grundmann, and N. Ledentsov, 
{\em Quantum Dot Heterostructures}\/ (John Wiley, New York, 1998).

\bibitem{zanardi:98}
P. Zanardi and F. Rossi, Phys. Rev. Lett. {\bf 81}, 4752 (1998).

\bibitem{quiroga:99}
L. Quiroga and N.F. Johnson, Phys. Rev. Lett. {\bf 83}, 2270 (1999).
     
\bibitem{motohisa:98}
J.~Motohisa, J.J.~Baumberg, A.P.~Heberle, J.~Allam,
Solid-State Electronics {\bf 42}, 1335 (1998).

\bibitem{landin:98}
L.~Landin, M.S.~Miller, M.E.~Pistol, C.E.~Pryor, L.~Samuelson,
Science {\bf 280} 262 (1998).

\bibitem{dekel:98}
E.~Dekel, D.~Gershoni, E.~Ehrenfreund, D.~Spektor,
J.M.~Garcia, M.~Petroff,
Phys. Rev. Lett. {\bf 80}, 4991 (1998).

\bibitem{zrenner:98}
A. Zrenner et al., Physica B {\bf 256--258}, 300 (1998).

\bibitem{hohenester:99a}
U. Hohenester, F. Rossi, and E. Molinari,
Solid-State Commun. {\bf 111}, 187 (1999).

\bibitem{hohenester:00} 
A.~Hartmann, Y.~Ducommun, E.~Kapon, U.~Hohenester, and E.~Molinari, 
\prl, in press (cond-mat/0005189);
R.~Rinaldi, S.~Antonaci, M.~DeVittorio, R.~Cingolani, U.~Hohenester,
E.~Molinari, H.~Lipsanen, and J.~Tulkki, \prb, accepted.
     
\bibitem{hawrylak:99}
P. Hawrylak, Phys. Rev. B {\bf 60}, 5597 (1999).

\bibitem{rinaldi:96}
R.~Rinaldi, P. V. Giugno, R. Cingolani,
H. Lipsanen, M. Sopanen, J. Tulkki, and J. Ahopelto,
Phys. Rev. Lett. {\bf 77}, 342 (1996);
Phys. Rev. B {\bf 57}, 9763 (1998).

\bibitem{dot} In our cylindrical QD the confinement energies due to the
in-plane parabolic potential are $\omega_o^{(e)}=20$ meV for electrons,
and $\omega_o^{(h)}=3.5$ meV for holes; with this choice, electron and
hole wavefunctions have the same lateral extension. The quantum-well
confinement along $z$ is such that the intersubband splittings are much
larger than $\omega_o^{(e,h)}$. Material parameters for GaAs are used.

\bibitem{rossi:96}
F.~Rossi, E.~Molinari, Phys. Rev. Lett. {\bf 76}, 3642 (1996);
Phys. Rev. B {\bf 53}, 16 462 (1996).

\bibitem{absorption} The optical absorption spectra are calculated in
linear response from the imaginary part of the Fourier-transformed
total interband polarization (for details see, e.g.,
Refs.~\onlinecite{rossi:96,hawrylak:98}). For conceptual clarity, in
Fig.~2 we have introduced a rather large broadening of $\sim 0.2$ meV
accounting for the finite lifetime of the excitonic states through
environment coupling (e.g., phonons).

\bibitem{hohenester:99b}
U. Hohenester, R. Di Felice, E. Molinari, and F. Rossi,
Appl. Phys. Lett. {\bf 75}, 3449 (1999).

\bibitem{remark:commensurable} Note that in a scenario where the
superposition of states is manipulated and exploited, as requested by
quantum algorithms, the phase changes due to the unperturbed
Hamiltonian play an important role and unavoidably have to be taken
into account. In this respect, external electric or magnetic fields
could be used for a fine-tuning of frequencies. It is worth mentioning
that the problem of keeping track of the unperturbed time evolution is
even more cumbersome in the case of NMR schemes, where the unperturbed
Hamiltonian also contains terms that couple the different qubits; see,
e.g., A. Steane., Rep. Prog. Phys. {\bf 61}, 117 (1998).


\end{thebibliography}
\end{document}